\documentclass[preprint]{emulateapj}
\newcommand{\etal}{et al.\ }
\newcommand{\beq}{\begin{equation}}
\newcommand{\eeq}{\end{equation}}

\newcommand{\msol}{{\rm M_{\odot}}}

\begin{document}

\title{Recent OH Zeeman Observations: \\Do they Really Contradict the 
Ambipolar-Diffusion Theory of Star Formation?}
%

\author{Telemachos Ch. Mouschovias\altaffilmark{1} and Konstantinos Tassis\altaffilmark{2,3}}
\altaffiltext{1}{Departments of Physics and Astronomy, University of Illinois at 
Urbana-Champaign, 1002 West Green Street, Urbana, IL 61801}
\altaffiltext{2}{Department of Astronomy and Astrophysics and Kavli Institute of 
Cosmological Physics, Chicago, IL 60637}
\altaffiltext{3}{Current Address: Jet Propulsion Laboratory, California Institute of 
Technology, Pasadena, CA, 91109}

\begin{abstract}
Until recently, many of the dozens of quantitative predictions of the
ambipolar-diffusion theory of gravitational fragmentation (or core formation) 
of molecular clouds have been confirmed by observations and, just as importantly, 
no prediction has been contradicted by any observation. A recent paper, however, 
claims that measurements of the variation of the mass-to-flux ratio from envelopes 
to cores in four clouds {\it de}creases, in direct contrast to a prediction of the 
theory but in agreement with turbulent fragmentation (in the absence of gravity) 
and, therefore, the ambipolar-diffusion theory is invalid (Crutcher \etal 2008). 
The paper treats magnetic-field nondetections as if they were detections. 
We show that the analysis of the data is fundamentally flawed and, moreover, 
the comparison with the theoretical prediction ignores major geometrical effects, 
suggested by the data themselves if taken at face value. The magnetic fluxes of 
the envelopes are also miscalculated. We carry out a proper error analysis and 
treatment of the nondetections and we show that the claimed measurement of the 
variation of the mass-to-flux ratio from envelopes to cores is not valid, no 
contradiction with the ambipolar-diffusion theory can be concluded, and no 
theory can be tested on the basis of these data.
\end{abstract}


\maketitle

\section{Introduction}\label{intro}

The ratio of the mass and magnetic flux of interstellar molecular clouds 
has received well-deserved observational attention in recent years (e.g., 
Crutcher 1999; Heiles \& Cructher 2005). For a cloud as a whole, the 
mass-to-flux ratio is important input to the ambipolar-diffusion theory 
of fragmentation (or core formation) in molecular clouds. For fragments 
(or cores) within a cloud, the ambipolar-diffusion theory makes specific 
predictions of the mass-to-flux ratio and its evolution in time. Until now, 
observations have been in excellent quantitative agreement with the 
theoretical predictions (e.g., see Crutcher \etal 1994; Crutcher 1999 and 
correction by Shu \etal 1999, pp. 196 - 198; Ciolek \& Basu 2000). Recently, 
however, Crutcher \etal (2008) (hereinafter CHT) have looked at the variation 
of the mass-to-flux ratio from the envelopes to the cores of four molecular 
clouds and concluded that there is a fundamental discrepancy between the 
observations and the predictions of the ambipolar-diffusion theory.

In what follows, we examine critically the claimed discrepancy. Before we 
undertake a direct re-examination of the data and the conclusions drawn 
from them, it is important that we summarize the key elements of the 
ambipolar-diffusion theory of protostar formation. This is especially 
necessary because, often, what is a prediction of the theory is quoted as 
if it were a requirement or assumption, and vice versa, i.e., input of the 
theory taken directly from observations is often quoted as if it were a 
prediction. The ambipolar-diffusion timescale is also often a frequent 
victim of abuse in the astronomical literature. In this context, it is  
of utmost importance to clarify the difference in the way the process of 
ambipolar diffusion affects the ordered (or mean) component of the magnetic 
field in a molecular cloud and the magnetohydrodynamic (MHD) waves (or 
turbulence) that are superimposed on the mean field. It is also useful, 
for putting the current controversy in its proper context and perspective, 
to summarize how theory and observations of interstellar magnetic fields 
have interacted over the years and what the outcome of that interaction has 
been.

We summarize in \S \ref{theory} the main elements of the ambipolar-diffusion 
theory in a context relevant to the present controversy and to maintaining 
proper perspective. In \S \ref{zeeman} we first summarize the observational 
procedure followed by CHT and state their main conclusion, and then we state
three major flaws in their analysis as well as three additional such flaws in 
their comparison of the results with the ambipolar-diffusion theory.

In \S \ref{A}, we explain the proper procedure for combining multiple observations 
of the envelope magnetic field in each cloud to derive an estimate of the mean and 
an associated uncertainty. An explicit assumption of the analysis by CHT is 
that line-of-sight effects (as encoded in $\cos \theta$) are the same for the core and 
the envelope. An implicit assumption of the same analysis is that the magnetic field 
does not vary intrinsically from one envelope location to another. None of these assumptions 
is justified by the observations. Relaxing them in a manner suggested by the data themselves 
alters the manner in which errors should be treated and increases significantly the 
uncertainties of all results. 

In \S \ref{B}, we discuss the difference between a detection and an upper limit of the 
mean envelope magnetic field. Most of the individual observations yielded nondetections, 
a fact recognized by the authors. The authors state that for this experiment detections 
are not necessary. No upper limits are calculated, which would be appropriate for 
nondetections; instead, nondetections and detections are treated in the same way with respect 
to error propagation and handling, as well as for quoting results. We remedy this problem and 
explicitly calculate upper limits where appropriate, using a full Monte Carlo analysis of 
the error propagation to properly account for the (large) errors propagating through a 
nonlinear equation (see eq. [\ref{ther}] below). 

In \S \ref{C}, we elaborate certain theoretical issues (points 3a, b, c of \S \ref{zeeman}) 
which should be paid due consideration if a meaningful comparison between theory 
and observations is to be achieved at least as far as the variation of the mass-to-flux 
ratio from the envelope to the core of a molecular cloud is concerned.

\section{The Development of the Ambipolar-Diffusion Theory, its Main Elements, and
Recent OH Zeeman Observations}\label{ADandZeeman}

\subsection{Key Elements of the Ambipolar-Diffusion Theory of Protostar Formation}
\label{theory}

Ambipolar diffusion (the motion of the two poles of charge, ions and electrons and 
possibly charged grains, which at densities characteristic of molecular clouds are well 
attached to magnetic field lines, relative to the neutral particles) is an unavoidable 
process in weakly ionized, self-gravitating systems, e.g., molecular clouds. It occurs 
on a timescale given by (see Mouschovias 1979b)
\beq\label{TAD}
\tau_{\rm AD} = 1.8 \times 10^{6} \, \left( \frac{x_{\rm i}}{10^{-7}} \right) \, \, \, {\rm yr},
\eeq
where $x_{\rm i}$ is the degree of ionization. Unlike the early understanding of this 
process, which envisioned magnetic flux escaping a cloud as a whole and thereby depriving 
a cloud of magnetic support and allowing it to collapse and form stars (e.g., Mestel \& 
Spitzer 1956; Nakano \& Tademaru 1972; Nakano 1973, 1976, 1977), a re-examination 
of the process concluded that ambipolar diffusion is much more effective in the deep 
interiors of molecular clouds, where the degree of ionization is smaller than $10^{-7}$, 
than in their envelopes, where the degree of ionization is usually greater or much 
greater than $10^{-5}$. The important difference in the two pictures is that, in the latter, 
a cloud as a whole does not lose any magnetic flux. Instead, neutral particles, under 
their own self-gravity, contract slightly more rapidly than the plasma (and the field
lines), thereby redistributing the mass in the central parts of a dense cloud and 
increasing the mass-to-flux ratio in the central flux tubes (but not necessarily in 
an individual fragment or core; see below).

Early efforts to understand the possible role of ambipolar diffusion in cloud dynamics 
and star formation consisted almost exclusively of comparing $\tau_{\rm AD}$ (actually, 
an earlier version of this timescale obtained by Spitzer 1968 for uniform, cylindrical 
cloud models and which was greater than that given by eq. [1] by almost a factor of 3) 
and the free-fall time, which for a spherical cloud model of uniform neutral density 
$\rho_{\rm n}$ is 
given by
\beq
\tau_{\rm ff} = \left(\frac{3 \pi}{32 G \rho_{\rm n}}\right)^{1/2} \,
 = 1.07 \times 10^{6} \, {\left(\frac{10^{3} \, {\rm cm^{-3}}}{n_{\rm n}}\right)}^{1/2} 
\, \, \, {\rm yr} .
\eeq
Since $\tau_{\rm AD}$ was invariably found to be greater than $\tau_{\rm ff}$, 
it was always concluded that ambipolar diffusion is not a relevant process in star 
formation. The presumption was that a cloud gives birth to stars on the free-fall 
timescale, as orginally suggested by Hoyle in his hierarchical fragmentation picture, 
or on a somewhat longer ("`magnetically diluted") free-fall time because of the 
presence of a magnetic field frozen in the matter (Mestel 1965).

A series of papers initiated and developed the main elements of the modern theory  
of single-stage fragmentation (or, core formation) and star formation in molecular clouds 
as a natural consequence of primarily magnetic support of such clouds (Mouschovias 1977,
1978, 1979, 1981, 1987a,b, 1989, 1991a, b, c; Shu \etal 1987). Observations were already 
suggesting that star formation does not occur on the free-fall timescale but on a timescale 
exceeding $10^{7}$ yr (e.g., see Roberts 1969; Zuckerman \& Palmer 1974). Magnetic support was 
neither an arbitrary theoretical assumption nor an ingenious theoretical conclusion. It was a 
simple consequence of the fact that the then observed interstellar, large-scale magnetic field 
of 3 $\mu$G can easily support a mass of $5 \times 10^{5} \, \msol$ at the approximate mean 
density of the interstellar medium of $1\, {\rm cm^{-3}}$. Moreover, atomic-hydrogen clouds, 
thought to be the progenitors of molecular clouds have never been observed to have mass-to-flux 
ratios exceeding the critical value for collapse
\beq\label{mtfcrit}
\left(\frac{M}{\Phi_{B}}\right)_{\rm crit} = 
\left(\frac{1}{63G}\right)^{1/2},
\eeq
obtained from nonlinear, exact equilibrium calculations (Mouschovias \& Spitzer 1976). 
(The quantity $G$ is the universal gravitational constant.) 
In addition, if clouds were to begin their lifetimes in a magnetically supercritical 
state, contraction (ordered) velocities characteristic of collapse (several ${\rm km \, 
s^{-1}}$) would be commonplace. Instead, the gravitational potential energy of typical 
HI clouds is smaller than their magnetic energy by two orders of magnitude, and velocities 
characteristic of collapse for a cloud as a whole are not observed even for molecular 
clouds.

Mouschovias (1987a) described a new theory of star formation in primarily magnetically 
supported molecular clouds, with hydromagnetic waves (or turbulence) contributing part of 
the support against gravity. (At the time, it was referred to as a ``scenario'' because 
the rigorous, quantitative calculations needed to justify the term ``theory'', which 
implies predictive power, had not yet been undertaken.) Ambipolar diffusion, which as 
explained above is an unavoidable process in a weakly ionized, self-gravitating physical 
system such as a molecular cloud, acts differently on the large-scale (or mean) magnetic 
field and on short-wavelength disturbances associated with MHD waves or turbulence. 
Because of ambipolar diffusion, disturbances of wavelength $\lambda$ smaller than the 
{\it Alfv\'{e}n lengthscale}
\beq
\lambda_{\rm A} = \pi v_{\rm A} \tau_{\rm ni}
\eeq
cannot be sustained in the neutrals; i.e., they decay because of neutral-ion friction. 
(The quantity $v_{\rm A} = B/(4\pi \rho_{\rm n})^{1/2}$ is the Alfv\'{e}n speed in 
the neutrals, and $\tau_{\rm ni}$ the mean (elastic) collision time of a neutral 
particle in a sea of ions.) Unlike the timescale for redistribution of mass in the 
central flux tube of a cloud given by equation (\ref{TAD}), the timescale for the decay of 
such disturbances is wavelength ($\lambda$) dependent and is given by
\beq\label{TADwave}
\tau_{\rm{AD,wave}} \approx 10^{5} 
\left(\frac{\lambda}{\rm{1 \, pc}}\right)^{2} \!\!\!
\left(\frac{30 \, \mu \rm{G}}{B}\right)^{2} \!\!\! 
\left(\frac{n_{\rm{n}}}{10^{3} \, {\rm cm^{-3}}}\right)\!\!\!
\left(\frac{x_{\rm i}}{10^{-7}}\right) \, \, \, {\rm yr}.
\eeq
This is a very short timescale even relative to free-fall, and it becomes smaller 
the smaller the wavelengths one considers. For typical molecular-cloud parameters, the 
Alfv\'{e}n lengthscale is arithmetically equal (0.3 pc) to the thermal critical 
lengthscale (essentially the Bonnor-Ebert radius for collapse of an isothermal sphere 
supported by thermal pressure against its self-gravity),
\beq
\lambda_{\rm{T,crit}} = 1.09 C_{\rm a} \tau_{\rm ff}
= 0.29 \left(\frac{T}{10 \, \rm{K}}\right)^{1/2} \!\!\!
\left(\frac{10^{3} \, {\rm cm^{-3}}}{n_{\rm n}}\right)^{1/2} 
 \, \, \, {\rm pc},
\eeq
where $C_{\rm a}$ is the adiabatic speed of sound in the gas.
Thus, if MHD waves (or turbulence) contribute even a small fraction of the support 
against gravity (e.g., a few percent), ambipolar diffusion removes that support very 
quickly and gravity remains locally unbalanced. Ambipolar diffusion, then, allows 
thermally supercritical but magnetically subcritical masses to begin to separate out as 
fragments (or cores). Eventually, the mass-to-flux ratio of the central flux tube 
of a cloud exceeds the critical value for collapse and dynamical contraction (but 
not free fall) ensues. This was the conceptual origin of the new theory of fragmentation 
and star formation suggested by Mouschovias (1987a; see also 1991a) and put on a 
rigorous footing in subsequent years (e.g., see review by Mouschovias 1996 and 
references therein). 

Statements frequently encountered in the literature, to the effect that 
the ambipolar-diffusion theory ignores MHD waves (or turbulence), are thus not based 
on the record. It is precisely because of the partial support that such disturbances 
provide against gravity and their consequent decay by ambipolar diffusion that 
fragmentation (or core formation) is initiated in this theory. It is noteworthy that, 
the greater the initial contribution of such disturbances to the support against 
gravity, the more rapid the ambipolar-diffusion--initiated fragmentation process 
becomes.

An important prediction of the detailed numerical simulations was that the central 
mass-to-flux ratio, after it exceeds the critical value given by equation (\ref{mtfcrit}) and 
dynamical contraction (but not free fall) sets in, does not continue to increase 
indefinitely. It asymptotes to a value typically 2 - 3 times greater than the critical 
one (e.g., Fiedler \& Mouschovias 1993, Fig. 9b; Ciolek \& Mouschovias 1994, Figs. 2e and 
4e). The physical reason for this is that the contraction is rapid (the infall acceleration 
is typically 30\% that of gravity), so that the magnetic flux is essentially trapped 
inside the contracting fragment. Once such gravitational contraction begins, unlike 
the early, ambipolar-diffusion--controlled phase of core formation in which the 
magnitude of the magnetic field does not typically increase by more than 30\%, 
the magnetic field increases with gas density as $B \propto \rho^{\kappa}$ with 
$\kappa \approx 1/2$.

Critics of the ambipolar-diffusion theory often refer to the 
large set of calculations that developed and refined the theory as ``static''. Far 
from that being the case, these calculations have demonstrated that, {\it {even if}} 
one begins with a model cloud that is in an initial exact equilibrium state and 
would therefore remain in such a state indefinitely if it were not for the presence of 
ambipolar diffusion, the fragmentation is relatively rapid (more precisely, it occurs on 
a timescale equal to 1/2 the value given by eq. [1]) and, moreover, these 
calculations are the only ones that have followed the subsequent {\it dynamical} phase of 
protostar formation, up to densities of about $10^{15} \, {\rm cm^{-3}}$, accounting 
for the zoo of phenomena relevant to the seven-fluid physical system (neutral particles, 
atomic and molecular ions, electrons, negatively-charged, positively-charged and 
neutral grains) -- e.g., see Tassis and Mouschovias (2007a, b, c). Mindful of the 
possibility that molecular clouds are not quiescent objects and are not necessarily 
found in quiescent environments, Mouschovias (1987a, fig. on p. 486; or 1991a, Fig. 1) 
considered both subAlfv\'{e}nic as well as superAlfv\'{e}nic contraction. 
Yet in both cases, ambipolar diffusion (and magnetic braking) plays a crucial 
role in protostar formation.

It was recognized early on in the development of the ambipolar-diffusion theory 
of fragmentation that, once a flux tube in a molecular cloud becomes magnetically 
supercritical, it may break up along its length into thermally supercritical but 
magnetically {\it sub}critical fragments (Mouschovias 1991c, \S 2.4). In fact, the 
factor by which the density of such fragments must increase in order for balance 
of forces along field lines to be re-established {\em and} a critical mass-to-flux 
ratio to be re-acquired was calculated analytically. Moreover, the effect that such 
fragmentation has on the $B - \rho$ relation was also determined. It is therefore 
of utmost importance to keep in mind these effects if one desires to make a 
statement about the mass-to-flux ratio of a single fragment relative to that of 
the parent cloud. In other words, if a magnetically critical or supercritical flux
tube breaks up along its length, the entire mass in the flux tube must be counted 
if one desires to obtain the relevant mass-to-flux ratio to compare\textit{} with the
theoretical prediction.

The theory has three significant dimensionless free parameters, whose typical values 
are obtained from observations (Fiedler \& Mouschovias 1992). Accounting for rotation 
introduces a fourth free parameter, which is essentially a measure of the moment of 
inertia of the cloud (or fragment) relative to that of a region in the surrounding 
medium (or envelope) of volume comparable to that of the cloud (or fragment) (Basu 
\& Mouschovias 1994). A simulation is carried out using those typical values of the 
free parameters as input. The calculation predicts the spatial dependence and time 
evolution of the physical quantities of the forming protostellar fragment and of the 
parent cloud (e.g., profiles of the density, magnetic field, angular velocity, infall 
velocity as functions of time). In the calculations without rotation, the three 
free parameters are: (1) the initial central 
mass-to-flux ratio $\mu_{\rm c0}$ in units of its critical value for collape; 
(2) the ratio of the initial free-fall time and the neutral-ion collision time, 
$\nu_{\rm{}ff,0}$; and (3) the exponent $k$ in the parametrization of the ion density 
$n_{\rm i}$ in terms of the neutral density $n_{\rm n}$, $k = d{\rm ln}{n_{\rm i}}/
d{\rm ln}{n_{\rm n}}$. If detailed grain chemistry is used to determine the degree 
of ionization at each stage of the evolution, the parameter $k$ is replaced by a quantity 
which is essentially the radius of a typical grain particle (Ciolek \& Mouschovias 1993). 
Once the results from the fiducial run are at hand, a parameter study is carried out. It 
consists of at least two additional runs for {\it each} of the free parameters, using 
values at the endpoints of the range suggested or allowed by observational uncertainties. 
Since the mass-to-flux ratio of the parent cloud (or envelope) is a dimensionless free 
parameter in the theory, the theory neither requires nor predicts that this quantity 
have any specific value, e.g., be subcritical.
The fact that, in any cloud in which a positive 
detection of the magnetic field component along the line of sight using Zeeman 
observations has been obtained, implies that the cloud's mass-to-flux ratio 
(given the geometrical uncertainties in estimating it) is consistent with a 
subcritical value is just that, observational input to the theory, not a requirement
or a prediction of the theory.

Early observational work using the Zeeman effect in the 21-cm line of HI, focused 
on testing the relation between the magnetic field strength and the gas density 
predicted by spherical, isotropic contraction, i.e., $B \propto \rho^{2/3}$ (e.g., 
Verschuur 1970, 1971). It was concluded that the observations were consistent with 
that relation. However, a more careful examination of the data showed that that 
was not the case (Mouschovias 1978, Fig. 1). Instead, the data was consistent 
with the prediction of the early, self-consistent calculations of the equilibrium 
and contraction of self-gravitating, isothermal, magnetic clouds embedded in a 
hot and tenuous external medium. Subsequent observations of the Zeeman effect in 
OH (which samples greater densities than HI) and in OH and $\rm{H_{2}O}$ masers 
(which sample even greater densities and reveal physical conditions in regions of 
active star formation) were found to be as predicted by the magnetic calculations 
(Fiebig \& Guesten 1989; Mouschovias 1996, Fig. 1b; Crutcher 1999).

More recently, observational attention has shifted from the $B - \rho$ relation 
to the mass-to-flux ratio, which, for a cloud as a whole is invaluable input to 
the ambipolar diffusion theory and, for fragments (or cores) can potentially be 
an important test of the theoretical prediction that ambipolar diffusion leads  
to an increase of the mass-to-flux ratio of a cloud's central flux tubes relative 
to the mass-to-flux ratio of the envelope.

\subsection {Recent {\rm OH} Zeeman Observations}\label{zeeman}

CHT have recently combined existing detections of the OH Zeeman 
effect in four molecular cloud cores with new observations of the same effect in 
the regions surrounding these cores, in an effort to get a handle on the variation of 
the mass-to-flux ratio from the envelope to the core of each cloud. The authors regard 
this as a definitive test of the ambipolar diffusion theory. The four clouds are 
L1448, B217-2, L1544, and B1. In the region 
surrounding each core in each of the four clouds, new observations were undertaken at 
four locations about each core (i.e., in the envelope of each core). At some locations 
in each envelope, positive detections were made while at others no detection could be 
achieved. For each envelope, the authors average the algebraic values of the four 
measurements of the line-of-sight magnetic field, regardless of whether they were 
detections or nondetections, and assign the resulting value to the average magnetic 
field strength of the envelope -- the assumption made here is that this value 
appropriately characterizes the mean magnetic field of the envelope. Using this field, 
CHT proceed to calculate what they regard as the magnetic flux of the 
envelope, which, combined with the flux in the core, is used to obtain the quantity 
$R$ defined by 
\begin{equation}\label{ther}
R=\frac{I_{\rm core} \Delta V_{\rm core}/{B_{\rm core}}} 
{I_{\rm env}\Delta V_{\rm env}/{B_{\rm env}}}.
\end{equation}
(The quantity $I$ is the peak intensity of the spectral line in degrees K, $\Delta V$ 
is the FWHM in km $\rm s^{-1}$, and $B$ is the line-of-sight magnetic field strength 
in microgauss.) They then average again the values of $R$ over the four clouds, and 
obtain an overall average value of $R$. The claim in the paper is that the derived 
value of $R$ is 5$\sigma$ away from 1 and that this contradicts a prediction of the 
ambipolar diffusion theory and, therefore, invalidates the theory, since the latter 
(according to CHT) predicts $R>1$.

We show quantitatively below that this claim, as stated by CHT, is not 
supported by their data; it is an artifact of the analysis and interpretation of the 
data, which suffers in three major ways: 

1. The treatment of the propagation of observational uncertainties systematically 
underestimates the uncertainties of the combined result at every step.

2. Nondetections of magnetic fields are treated as if they were detections, and upper 
limits are not quoted as appropriate -- in fact, based on the raw data presented in 
CHT, measurements of $R$ are not achievable in any of the clouds studied; 
rather, upper limits should have been calculated and quoted.

3. The comparison with ambipolar-diffusion calculations is itself flawed in three 
important respects: 

     (a) The theoretical simulations chosen for this comparison use as
     input a very different geometry of the field lines threading the cloud from the 
     geometry of the field lines that the observations presumably reveal. 

     (b) The magnetic flux threading each cloud is incorrectly calculated by taking 
     the arithmetic average of the four {\it algebraic} values of $B$ in each envelope. 
     
     (c) The possible breakup of the mass in the central flux tube of each cloud into 
     thermally supercritical but magnetically subcritical fragments (or cores), an 
     important possible effect in the theory (Mouschovias 1991c), is mentioned by the 
     authors but not taken into consideration in their analysis and/or arguments. 
     
If these effects are considered, it follows that a variety of $R-$values are 
possible. We discuss each of these points below.

\section{Combining Uncertainties and Propagating Errors}\label{A}

CHT quote values for the results of observations of the magnetic field at four different 
positions in the envelope of each cloud, together with associated uncertainties, derived 
as in Troland \& Crutcher (2008). In most cases, the quoted values correspond to 
nondetections. We discuss this issue in detail in the next section; here, we refer to 
detections and nondetections alike as ``observations''. 

For each observation of the envelope's line-of-sight magnetic field $B_j$, CHT quote an 
associated uncertainty $\sigma_j$, which we assume to be Gaussian. These values are given 
for each cloud in Table \ref{crutable}. The authors then calculate a mean value of the 
line-of-sight magnetic field in the envelope of each cloud, 
\begin{equation}\label{av}
B_{\rm mean} = \frac{1}{4}\sum_{j=1}^4B_j\,,
\end{equation}
and an uncertainty on the mean, $\sigma_{\rm mean}$, using Gauss' formula of error propagation 
on equation (\ref{av}), which gives 
\begin{equation}\label{gauss}
\sigma_{\rm mean} = \frac{1}{4}\sqrt{\sum_{j=1}^4 \sigma_j^2}\,.
\end{equation}
The results derived using this formula are given in the second column of Table 
\ref{statable}; they are the same as those quoted in CHT (see their Table 1, col. 4). 
\begin{table}
\begin{center}
\caption{\label{crutable} Magnetic Fields and Errors (in microgauss) in Four Cloud 
Envelopes (data from CHT).}
\begin{tabular}{c|rrrr}
\hline \hline 
Cloud & $B_1\pm\sigma_1$ & $B_2\pm\sigma_2 $& $B_3\pm\sigma_3 $ & $B_4\pm\sigma_4$ \\
\hline
L1448CO & $-9\pm 13$ & $-11\pm 6$ & $-7 \pm 7$ & $14\pm 8$ \\
B217-2 & $-13\pm 9$ & $5 \pm 6$ & $6 \pm 8$ & $9 \pm 13 $ \\
L1544 & $-3 \pm 4$ & $-1 \pm 4$ & $22 \pm 6$ & $2 \pm 10$ \\
B1 & $-16 \pm 6$ & $0 \pm 7$ & $-3 \pm 6$ & $-10 \pm 5$\\ 
\hline \hline
\end{tabular}
\end{center}
\end{table}
This analysis is a standard treatment of errors when combining several measurements 
{\em of the same quantity}. If CHT had mesured the magnetic field {\em at the same 
envelope location} in four different instances, their measurements could be combined 
using this procedure (although, even in this case, a {\em weighted average} and its 
associated error propagation formula would be more appropriate since the measurement 
uncertainties are not equal for every observation). Similarly, if there were an {\em 
a priori} reason to be certain that the envelope magnetic field is uniform and the 
four observations in each cloud are sampling the same value of the field, such 
treatment would be appropriate. However, this is not the case here, and the treatment 
of errors in CHT yields misleading results. We discuss the reasons below. 

First, it cannot be overemphasized that Gauss' formula for error propagation, and 
hence equation (\ref{gauss}), is not universally applicable. There are two reasons 
for which applicability might fail. The first stems from the fact that Gauss' formula 
is based on a Taylor expansion (e.g., see, Wall \& Jenkins 2003). As such, it is valid 
only if the errors being propagated are small, {\em or} if the equation through which 
the errors are propagated is linear (and thus no second-order terms enter the 
calculation). This condition does indeed apply in this case: although the errors are 
by no means small, the average is a linear operation and no second-order terms enter 
the error propagation formula. The second reason for which the applicability of Gauss' 
formula might fail is if the observations being combined {\em are not measuring the 
same intrinsic quantity}, but rather a distribution of values. Then, one's ability 
to constrain the mean, {\em especially} in the small-number--statistics regime, is 
not simply dependent on one's ability to take individual 
measurements; the possibility of having finite distribution spreads needs to be 
explicitly accounted for. 

The magnetic field in the cloud envelope is not known {\em a 
priori}  to have a unique uniform value everywhere in the envelope. In fact, the data 
seem to suggest exactly the opposite (e.g., compare observations 1 and 2 in cloud L1544 
with observation 3 in the same cloud - see Table \ref{crutable}). 
Equation (\ref{gauss}) is therefore not the appropriate estimate for the 
uncertainty on the mean magnetic field for each cloud envelope. What should one use instead? 

The spread in the observations is the convolution of the observational uncertainty and 
the intrinsic spread of the distribution of magnetic field values across the envelope. 
The sample variance can give an estimate of this convolved spread; its square root (the 
sample standard deviation, an unbiased estimator of the population standard deviation 
$\sigma$) is given by 
\begin{equation}
\sigma_S = \sqrt{ \frac{1}{3}\sum_{i=1}^4 \left(B_j - B_{\rm mean}\right)^2}\,.
\end{equation}
It is well known (e.g., see Wall \& Jenkins 2003; Lyons 1992) that the average, given 
by equation (\ref{av}), is distributed as a Gaussian about the true value of the mean, 
with variance $\sigma^2/N$ where $N$ is the number of observations. In our case, $N=4$ 
and the standard deviation of the average, which is the uncertainty on the mean, is 
$\sigma/2$. For all clouds, the best-guess estimate of $\sigma$, which is $\sigma_S$, 
is $\geq$ twice the uncertainty quoted on the mean (note, for example, the case of L1544 
where the uncertainty quoted by CHT on the mean is $3 {\rm \, \mu G}$ while $\sigma_S/2 
= 6  {\rm \mu G}$). This is an {\em empirical} proof that the uncertainties quoted by 
CHT are underestimated, even if one were to ignore the conceptual point made above.

%
%

Even $\sigma_S/2$, however, underestimates the true uncertainty on the mean. The 
uncertainty on the mean is $\sigma/2$, not $\sigma_S/2$. The sample standard deviation 
$\sigma_S^2$ is only an {\em estimate} of the population variance $\sigma^2$, and is 
distributed as $\sigma^2\chi^2/(N-1)$, where $\chi^2$ is a $\chi$-square variable with 
$N-1$ degrees of freedom (e.g., see Wall \& Jenkins 2003). In practical terms, this 
means that, in the low-statistics regime we are considering, $\sigma_S$ most frequently 
underestimates $\sigma$, and in about $10\%$ of all instances does so by at least a 
factor of $2$. In the latter case, the appropriate uncertainty on the mean is closer 
to $\sigma_S$ rather than $\sigma_S/2$. How do we then take into account not only that 
there is a finite population spread, but also that this population spread is essentially 
unknown, due to our inability to constrain it very well with the very limited number 
of available observations? 

We must use an analysis that allows not only for finite spreads, but also for a variety 
of such spreads, as long as they are allowed by the data. Allowing for a range of finite 
spreads consistent with the data increases the range of mean values that can explain the 
data, and in this way increases the uncertainty on the mean. Conversely, ignoring the 
possibility that there is true variation of mostly unknown extent of the $B_{\rm los}$ 
from point to point in the envelope {\em artificially} decreases the uncertainty in the 
mean $B_{\rm los}$ value.

\begin{table}
\begin{center}
\caption{\label{statable} Derived mean envelope magnetic fields and uncertainties. Two left 
columns: CHT  treatment; third column: sample standard deviation; fourth column: 
 likelihood analysis results (this work).}
\begin{tabular}{c|clc}
\hline \hline 
Cloud & $B_{\rm mean} \pm \sigma_{\rm mean} $& $\sigma_S$ & $B_{{\rm max}\mathcal{L} }\pm 
\sigma_\mathcal{L}$\\
\hline\\
L1448CO& $-3\pm 4$ & 12 & $-4 ^{+9}_{-8}$\\
\hline \\
B217-2 & $+2 \pm 5$  & 10 & $+2 ^{+7}_{-7}$ \\
\hline \\
L1544 & $+5\pm 3$   & 12 &$+4 ^{+10}_{-8}$ \\
\hline \\
B1 & $-7 \pm  3$ & 7 & $-8^{+5}_{-5}$ \\
\hline \hline
\end{tabular}
\end{center}
\end{table}

The most straightforward way to correctly combine these observations while allowing for 
finite spatial spread of $B_{\rm los}$ is through a likelihood analysis (e.g., see Wall 
\& Jenkins 2003; Lyons 1992; Lee 2004). In a likelihood analysis, the data can be assumed 
to be derived from a parent distribution of values (in our case, we assume that values 
of  $B_{\rm los}$ in the envelope follow a Gaussian distribution with mean equal to $B_0$ 
and intrinsic spread $\sigma_0$). This distribution is then ``sampled'' with $N$ measurements 
$B_j$, each carrying a (Gaussian) uncertainty of measurement $\sigma_j$. >From the data, we 
then calculate the joint probability distribution of the parameters $B_0, \sigma_0$ which 
are consistent with observations (the {\em likelihood}). In our case, the likelihood can 
be shown to be (e.g., see Venters \& Pavlidou 2007, and appendix \ref{apx})
\begin{equation}\label{likelihood}
\mathcal{L}\left(B_{0}, \sigma_{0}\right) = 
\left(
\prod_{j=1}^{N}\frac{1}{\sqrt{\sigma_0^2+\sigma_{j}^2}}
\right)
\exp\left[-\frac{1}{2} \sum_{j=1}^{N}
\frac{(B_{j} - B_{0})^2 }{\sigma_0^2+\sigma_{j}^2}
\right]\,.
\end{equation}
Any parameters that are not of direct interest to us (such as $\sigma_0$ in our case), can 
then be integrated out of the likelihood. In this way, we can derive the probability 
distribution of the parameter of interest ($B_0$ in our case) {\em while still allowing for 
all possible values in $\sigma_0$}. The integrated likelihood is called the {\em marginalized 
likelihood}, $\mathcal{L}_m$; this probability distribution can then be used to derive 
confidence intervals and upper limits where appropriate. A detailed discussion of the 
likelihood approach to error analysis, and the use of the marginalized likelihood to 
derive confidence intervals and upper limits, is given in Appendix \ref{apx}. An example 
of the (unnormalized) marginalized likelihood for the case of L1448CO is shown in Figure 
\ref{marginalized}. 
The marginalized likelihood is derived by numerically integrating equation (\ref{likelihood}) 
over $\sigma_0$ for different values of $B_0$, and is shown as a solid line; the location of 
the maximum-likelihood estimate for the mean $B_0$ is indicated by the dash-dot line. 

The maximum-likelihood estimates and associated uncertainties of $B_0$ are shown in Table 
\ref{statable}. It is clear that these are systematically greater than the errors produced 
by (the nonapplicable in this case) equation (\ref{gauss}), and they are generally greater 
than $\sigma_S/2$. The value of $\mathcal{L}_m$defining the $1\sigma$ uncertainties in the 
case of L1448 and the associated $B_{0-}$ and $B_{0+}$ are marked with the dashed
lines in Figure \ref{marginalized}. For comparison we show, with the shaded box, 
the $1\sigma$ spread of the values of $B_0$ that CHT  quote for the same object, 
based on the same data. 

Besides the significant underestimate of uncertainties in the mean envelope magnetic 
field  quoted in CHT, the second striking feature seen in Figure \ref{marginalized} is 
that the ``measured'' mean envelope magnetic field is very much consistent with zero 
(even using  the CHT uncertainties, zero is marginally more than one $\sigma$ away from 
their ``best-guess'' mean), at least in the case of L1448CO. It is also clear from Table 
\ref{statable} that, the maximum-likelihood values of $B_0$ for {\em all} clouds are no more 
than $2\sigma$ away from zero. As a result, the observational data presented in CHT are 
inconsistent with a {\em mean} value of the envelope magnetic field equivalent to a 
``detection'' in {\em any} of the observed clouds (note however that in certain individual 
locations within cloud envelopes detections have been achieved; the significance of these 
results is nevertheless diluted by the rest of nondetections, and the mean is never more 
than $2\sigma$ away from zero). Given this fact, one would expect to see appropriate upper 
limits quoted for the mean magnetic field in the envelope for each of these clouds. Such 
limits are not given; instead, the ``best-guess'' values and their uncertainties are 
treated in exactly the same way that one should treat only detections. This makes the 
interpretation of the data presented in CHT problematic.
We discuss this issue in the following section. 

\section{Upper Limits {\it versus} Detections}\label{B}

A ``detection'' of a magnetic field is an observation with a high enough signal-to-noise 
ratio that the measured magnetic field is statistically significantly different from zero. 
In other words, when a detection occurs, we are confident at some level ($99.7\%$ for a 
nominal $3\sigma$ detection) that the magnetic field measurement is not spurious (a result 
of noise). If a detection has indeed occured, then the magnitude and sign of the magnetic 
field can be quoted, together with associated error bars (indicating the $1\sigma$ 
uncertainty in the magnetic field value). If a detection has not occured, a value of 
the magnetic field is meaningless and is therefore not quoted; what is routinely quoted 
instead is an {\em upper limit} (a value of the magnetic field magnitude which, if 
exceeded in the observed system, 
would lead to a detection). 

In fields of science such as particle physics detections and upper limits are routinely 
quoted at the $5\sigma$ level; even in the literature of Zeeman observations of interstellar 
magnetic fields it is common practice to adopt as detections $3\sigma$ measurements, and 
accordingly quote $3\sigma$ upper limits where appropriate (e.g. Crutcher \etal 1993). 
However, for consistency with CHT and what they  consider as a tentative detection, we 
adopt $2\sigma$ limits here. Formally, then, the $2\sigma$ upper limit on the magnetic 
field is the value of the magnetic field for which we are 
confident at the $95.4\%$ level that it is not exceeded in the observed system. 

\begin{figure}
\includegraphics[width=3in]{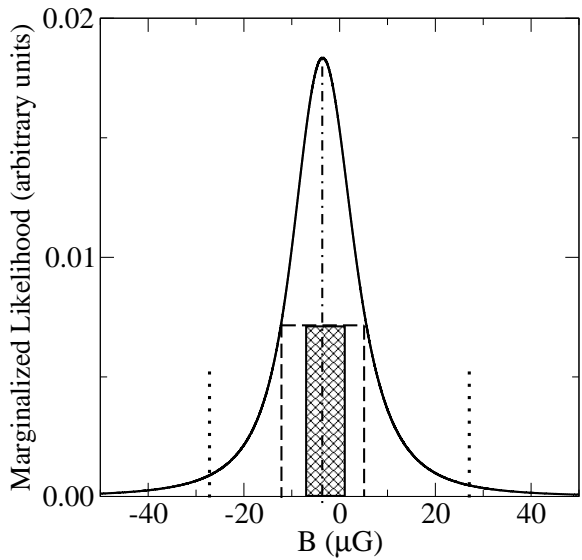}
\caption{Marginalized likelihood for the case of L1448CO (not normalized, heavy black line). 
The location of the maximum-likelihood value of $B_0$ is shown with the dot-dashed line. The 
dashed lines indicate the $1\sigma$ equilikelihood level, and the locations of the $1\sigma$ 
limits of $B_0$. The shaded box shows the allowed $1\sigma$ values for the mean 
envelope magnetic field for the same cloud according to CHT. Dotted lines indicate 
the $2\sigma$ upper limit on the absolute value of the mean envelope magnetic field. 
\label{marginalized}}
\end{figure}

Even if we use the (significatly underestimated, as we discussed in the previous section) 
overall uncertainties of CHT  quoted in Table \ref{statable}, and treat them as perfectly 
gaussian errors (which, as we have seen, is not the case), 3 of the 4 clouds do not have a 
statistically significant ($\geq 2\sigma$) measurement of the mean envelope magnetic field, 
while the 4th cloud (B1) has a marginal, $2\sigma$ measurement of the mean; even the B1 mean 
envelope magnetic field becomes consistent with zero if the correct uncertainties are used. 
As CHT point out, this is not necessarily a problem: useful limits on the ratio $R$ can be 
derived using upper limits for the envelope magnetic field, provided that measurements of 
the core magnetic field exist. However such limits, 
either for the envelope magnetic field or for $R$, {\em are not calculated by the authors}. 
Instead, nondetections are treated as detections, and the authors repeatedly refer to their 
``measurements'' of $R$, $R'$, and their mean. {\it Such measurements are not, in fact, 
possible with the data presented in their paper. The ratio R has not been measured}. However, 
upper limits can be placed on its value although, as we explain below, these limits are not 
very strong for any of the observed clouds. 

We note that the proper treatment of nondetections can in principle severely constrain 
theories of interstellar medium evolution and core formation, if the upper limits placed 
are strong enough (it could, for example, be the case that the measured $3\sigma$ upper 
limit for $R$ is $10^{-5}$! 
-- that would certainly be implying a very weak envelope magnetic field and a very puzzling 
situation for {\em any} current core formation theory). However, the magnetic field upper 
limits have to be correctly calculated and treated as such -- and they do have to be strong 
enough, which is not the case at all for the data presented in CHT. 

Upper limits for the (absolute value of the) mean magnitude of the line-of-sight magnetic 
field {\em combining all measurements of the envelope magnetic field} can be derived starting 
from the marginalized likelihood of the previous section. The procedure is discussed in detail 
in Appendix \ref{apx}. 
Utilizing the marginalized likelihood provides a very straightforward way to account for 
and correctly treat the asymmetric nature of the error bars in the combined observational 
results for the envelope magnetic field, the long non-gaussian tails, and the exact nature 
of the question we are trying to answer. 

The $2\sigma$ upper limits for the envelope magnetic field calculated 
this way are given in Table \ref{limtable}. The values of $|B_0|$ 
and $-|B_0|$ which include between them a fractional area of $2\sigma$ of the marginalized 
likelihood for L1448CO are marked with dotted lines in Figure \ref{marginalized}. Note that 
the marginalized likelihood at $-|B_0|$ and at $|B_0|$ need not have (and does not have in 
this case) the same value; this is frequently encountered when the robust definition of an 
absolute-value mean envelope $B$-field upper limit is used.

Since no measurements of the {\em mean} envelope magnetic field $|B_{\rm env}|$ have been 
achieved, no meaningful values of the ratio $R$ can be inferred. We can, however, use the 
derived upper limits to similarly place upper limits on the values of $R$. To treat the 
problem properly, and correctly propagate uncertainties on the measured or limited observables 
to the derived quantity $R$, we {\em cannot} use equation (\ref{gauss}), since the individual 
uncertainties involved are large while at the same time the equation which defines $R$ 
(equation [\ref{ther}]) is nonlinear. Instead, {\it we do a full Monte-Carlo calculation 
to properly derive the probability distribution for the values of R as follows}. We repeat 
the  following experiment $10^6$ times: we draw $I_{\rm core}$, $I_{\rm env}$, $\Delta 
V_{\rm core}$, $\Delta_{\rm core}$ and $B_{\rm core}$ from gaussian distributions with mean 
and spread equal to the measurement and uncertainty quoted in CHT (this is equivalent to 
assuming all errors in these quantities to be Gaussian); we draw a mean value of 
$B_{\rm env}$ from the marginalized likelihood of the previous section; we combine all 
the ``mock observations'' of these numbers to produce one value of $R$. We use the $10^6$ 
values of $R$ produced in this way to numerically calculate the probability distribution 
for $R$. We then calculate the $2\sigma$ upper limit on $|R|$ by requiring that the 
fractional integral of this distribution between $-R$ and $R$ be $95.4\%$. The $2\sigma$ 
upper limits for $|R|$ are given in Table \ref{limtable}.
Unfortunately, none of these limits are very strong. Although this is an interesting thought 
experiment, {\it much more and better quality data are required before any statement can be 
made with regard to comparison with any theory}. 

In the data presented by CHT, there is a single statistically significant, at the 
$3\sigma$ level, detection of the envelope magnetic field -- one of the four observations of 
the L1544 envelope magnetic field yields a value of $22\pm 6 {\, \rm \mu G}$. This measurement 
is consistent with the $29 \, {\rm \mu G}$ $2\sigma$ upper limit that one derives if 
all four observations of the envelope magnetic field are combined for the particular cloud. 
If we use {\em this} value with its associated uncertainty to calculate an $R$ for the case 
of L1544, we obtain $3.62 \pm 1.29$ -- {\it a marginal detection of the ratio R, with a value 
of a few}. Again, this value is consistent with the derived overall $3\sigma$ upper 
limit of $5.0$ obtained for L1544 by combining all observations of the magnetic field.

\begin{table}
\begin{center}
\caption{\label{limtable} $2\sigma$ Upper Limits on the Envelope Magnetic 
Field (in ${\rm \mu G}$) and on $R$.}
\begin{tabular}{c|cc}
\hline \hline 
Cloud & $|B_{\rm env}|(\leq 2\sigma)$ & $|R| (\leq 2\sigma)$  \\
&    $({\rm \mu G})$ &  \\
\hline
L1448CO & $27$ & $2.0$ \\
B217-2 & $22$  & $2.9$ \\
L1544 & $29$  & $5.0$  \\
B1 & $20$  & $1.1$  \\
\hline \hline
\end{tabular}
\end{center}
\end{table}

\section{Comparison with the Ambipolar-Diffusion Theory} \label{C}

\begin{figure}
\includegraphics[width=3in]{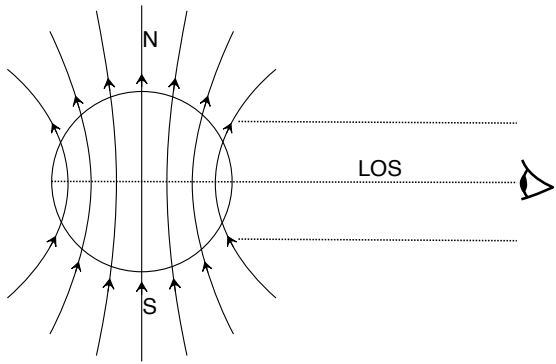}
\caption{ 
Schematic diagram of a star (e.g., Sun) that has a dipolar magnetic
field and is observed along lines of sight parallel to its
equatorial plane. To calculate the magnetic flux threading the
star, only the mean value of the magnetic field in either the
northern {\it or} in the southern hemisphere is needed. If both
values are averaged algebraically, the erroneous flux value of
zero will be obtained.
\label{starfield}}
\end{figure}

The main conclusion of the CHT paper is that their measured values of 
$R$ are significantly smaller than 1, which is in contradiction with a prediction 
of the ambipolar-diffusion theory. 
We have shown above that there are, in fact, no measurements of $R$ possible with the 
CHT data; only weak upper limits can be placed, which have values of a few. 
This, however, is only part of the problem with their strongly stated conclusions. 

The second error stems from the manner in which they calculate the magnetic flux of each 
cloud, even if their envelope magnetic fields and associated errors were to be exactly as 
quoted in their paper. They add algebraically the four values of the ``measured'' 
line-of-sight magnetic field strength and divide by four to obtain the mean field of the 
envelope. (Then they multiply this value by the area of each envelope {\it in the plane of 
the sky} to get the magnetic flux.) This procedure badly underestimates the envelope's 
magnetic flux in at least three of the four cases, in which the envelope ``measurements'' 
show fields with opposite algebraic signs at different positions. These reversals in the 
direction of the magnetic field of the envelope, if real, would imply a bent magnetic 
flux tube threading the cloud. In such a case, only one algebraic sign of the 
magnetic field (the one corresponding to the greatest
absolute values) should be considered in estimating the magnetic flux of the envelope. An
example will clarify the point. Suppose that the dominant component of the magnetic field
of a star, such as the Sun, is dipolar in nature and suppose that one wants to estimate 
the magnetic flux threading the star by measuring the magnetic field on the star's surface.
We assume, for simplicity, that the line-of-sight from the observer to the star lies in 
the star's equatorial plane, as shown in Figure \ref{starfield}. If this observer did what CHT
do, he would conclude that the magnetic flux of the star is exactly zero. That, of course, 
is far from being the case. The magnetic flux threading the star is obtained from the 
mean value of the magnetic field of either the sourthern or the northern hemisphere of 
the star (but not from the algebraic sum of the two) by multiplication with the area of 
the equatorial plane of the star.

\begin{figure}
\includegraphics[width=3in]{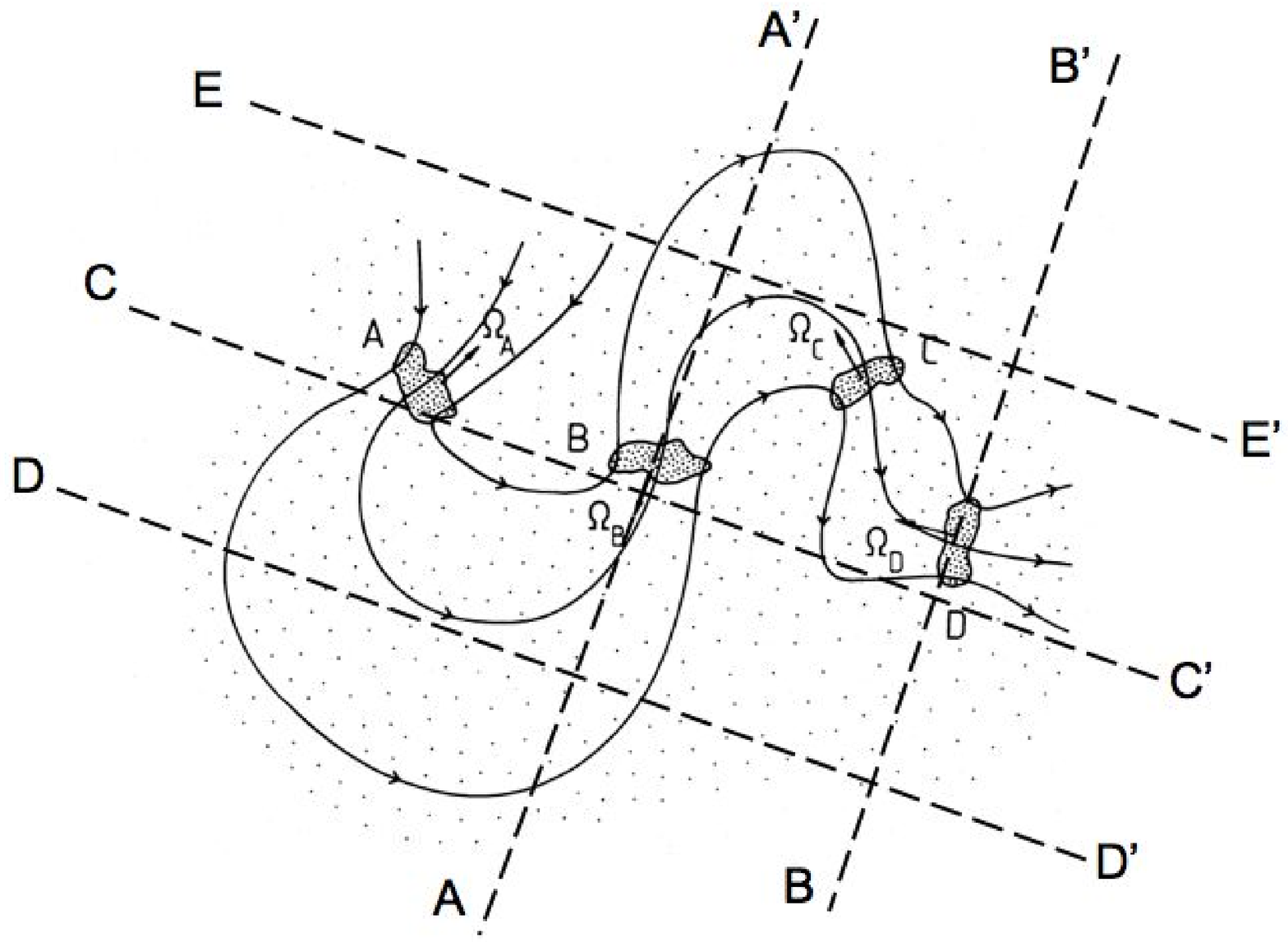}
\caption{\label{manycores}Schematic diagram (from Mouschovias and Morton 1985b) of a 
deformed flux tube that has fragmented along its length in 
a molecular cloud. The deformation can be caused by the relative 
motion of the fragments. The fragments need not be magnetically 
connected for their relative motion to cause significant 
deformations of the field lines. The dashed lines represent 
different lines of sight, whose significance is explained
in the text.}
\end{figure}

\begin{figure}
\includegraphics[width=3in]{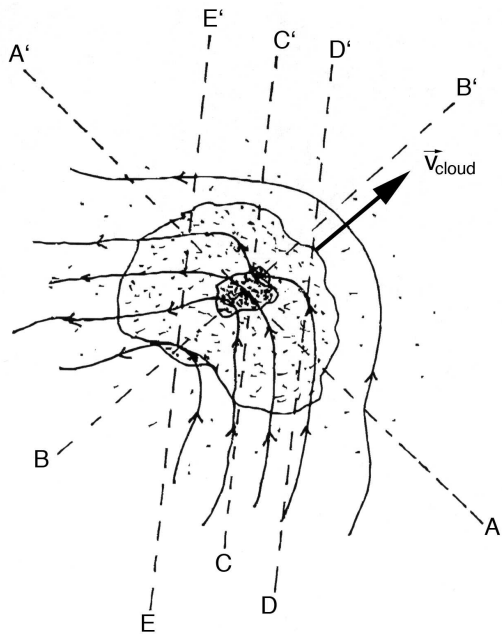}
\caption{ \label{loss}
Schematic diagram of the deformation of the field lines threading
a cloud caused by its motion relative to the surrounding medium.
The cloud is shown, for simplicity, to contain only one fragment
(or core), in the neighborhood of which the hourglass shape of
the field lines had been established during core formation but
affected by the cloud's motion. The dashed lines represent
different lines of sight, whose significance is explained in
the text.
\label{cartoon}}
\end{figure}

Moreover, although CHT believe that their data reveal reversals in the direction
of the magnetic field within the clouds they observe, they nevertheless compare their 
results with calculations that follow the formation and evolution of fragments in 
molecular clouds under the assumption that the parent clouds' magnetic field lines 
are initially essentially straight and parallel. This idealization in the theoretical 
calculations renders a mathematically complicated multifluid, nonideal MHD system tractable 
while capturing all the essential physics of the core formation and evolution problem. 
However, nobody has ever suggested that in a real cloud the magnetic field lines will be
essentially straight and parallel, except for the local hour-glass distortions associated 
with the compression of the field lines during gravitational core formation. In fact, 
Mouschovias \& Morton (1985b, Fig. 13) had sketched what they regarded as a more realistic
field geometry in a molecular cloud in which there are several (in that case four) magnetically
connected fragments. That figure is reproduced here as Figure \ref{manycores}. This configuration 
can result from relative motion of the fragments (labeled A, B, C, and D in the figure) within 
the cloud. The motion of a cloud as a whole relative to the intercloud medium will also 
bend the magnetic field lines in an almost U shape, as sketched in Figure 4. One can easily
visualize lines of sight in Figures \ref{manycores} and \ref{loss} along which, if one were 
to observe $B_{\rm los}$,
one would detect reversals and/or essentially vanishing $B_{\rm los}$. In other words, 
observations that can potentially reveal the geometry of the field lines can and should 
be used as input to make refinements in the theory and/or build a particular model for 
the cloud observed (as, for example, done in the case of B1 by Crutcher \etal 1994, and 
in the case of L1544 by Ciolek \& Basu 2000); such observations by no means invalidate 
the theory, which afterall has produced dozens of predictions many of which have been 
confirmed by observations and none of which has been contradicted by any observation.

Figures \ref{manycores} and \ref{loss} also demonstrate that the assumption by CHT, namely, that 
$\cos \theta$, the geometrical correction accounting for the fact that only the line-of-sight 
magnetic field is measurable by Zeeman observations, is the same in the core and everywhere 
in the envelope, is not in general valid. 

The lines-of-sight labeled AA', BB', CC', DD' and EE' in Figure 4 are chosen so as to 
reveal qualitatively similar behavior of the core and envelope $B_{\rm los}$ as the 
correspondingly labeled lines-of-sight in Figure \ref{manycores}. In relation to Figure 
\ref{loss}, if an observer's line-of-sight is the line DD', he'll detect essentially 
the full strength of the magnetic field in the nearest side of the cloud's envelope. 
Along CC', he will detect a weaker $B_{\rm los}$ in the core, although 
the full strength of the core's magnetic 
field is clearly greater than that of the envelope, as revealed by the compressed field
lines. By contrast, along the EE' line-of-sight no envelope $B_{\rm los}$ will be detected; 
and along AA', almost the full strength of the core's magnetic field will be measured, but
only a fraction of the envelope's magnetic field strength will be detected. It is clear 
from this illustrutive example, which actually may not be very different from the field 
line geometry in a real cloud (e.g., see Brogan \etal 1999), that unless one is mindful 
of geometric effects, incorrect conclusions will be reached on the true magnitude of the 
magnetic field and, therefore, of the mass-to-flux ratio in cloud cores and envelopes. 
When one recalls that the relevant column density for calculating the mass-to-flux ratio 
is the one {\it along a mangetic flux tube}, not the one measured along an arbitrary 
line-of-sight, which may be at a very large angle with respect to the true magnetic 
field direction, one realizes that geometric effects cannot be ignored in attempts to 
measure or estimate the mass-to-flux ratio of a cloud or a core (e.g., see Shu \etal 
1999). Simply taking the ratio of the two does not eliminate the need to account for 
the geometric effects.

\section{SUMMARY AND CONCLUSION}\label{disc}

We have examined critically the analysis and conclusions of recent OH Zeeman  
observations of four molecular cloud envelopes and cores by Crutcher \etal (2008), 
and we have shown that:
\begin{enumerate}
\item The error analysis of measurements of the envelope magnetic field is flawed, 
and the uncertainties on the observed mean value of the field are systematically 
underestimated in every cloud. 
\item Although the mean value of the envelope magnetic field is not statistically 
significantly different from zero in any cloud, the authors treat these values as 
if they were detections, and they do not give appropriate upper limits. 
\item If the error analysis is performed correctly, allowing possible spatial 
variations of the line-of-sight magnetic field (suggested by the data), and upper 
limits are appropriately calculated for the mean magnetic field in the envelopes, 
these upper limits are very weak: the mean magnetic field in the envelopes is 
constrained to be smaller than $20-30{\rm \, \mu G}$ at the $2\sigma$ significance 
level. 
\item If corresponding $2\sigma$ upper limits are calculated for the value of the 
ratio $R$, which is given by equation (7), they are also weak: $R$ is constrained 
to be smaller than a few. For no cloud is the upper limit smaller than $1$. $R$ 
is never statistically significantly different from zero, and has not been 
``measured'' for any cloud. 
\item The magnetic fluxes of the clouds were calculated incorrectly, by averaging 
{\it algebraically} the (presumed) values of the line-of-sight magnetic field 
$B_{\rm los}$ over a cloud's area projected on the plane of the sky, despite claimed 
reversals in the field direction not along one and the same line-of-sight, but 
from one line-of-sight to another. 
\item The possible gravitational breakup (of the mass {\it along} the flux tubes 
threading the cores) into thermally supercritical but magnetically subcritical 
fragments, an effect of the ambipolar-diffusion theory studied early on (Mouschovias 
1991c), was ignored by the authors. Although it is not clear whether this effect 
has occurred in the four clouds studied, it is nevertheless important to consider 
it properly before bold conclusions are drawn on the validitiy of the ambipolar-
diffusion theory. 
\item The ambipolar-diffusion theory can explain values of $R$ smaller than 1, 
especially if three-dimensional geometrical effects are properly accounted for. 
Although such effects are indeed missing from the published (axisymmetric) 
calculations against which CHT compare their observations, this is hardly new 
information. Focusing on effects that are known {\em a priori} to be missing from 
specific theoretical calculations is not a proper way to test the underlying 
theory or to make observational progress toward distinguishing between alternative 
theories. Such observations, when reliable, would provide useful {\it input} to 
the ambipolar-diffusion theory. 
\end{enumerate}

\appendix

\section{Likelihood Approach to Error Analysis}\label{apx}

In the classical sense, the likelihood represents the probability to have observed 
the data given some parent probability distribution. In the Bayesian sense, the 
likelihood is also proportional to the probability of the parent distribution to 
be described by a specific set of parameters (mean, variance) given the data. In 
a likelihood analysis such as the one presented here, the likelihood (properly 
normalized) {\em is} the probability of the set of parameters we are trying to 
estimate (mean, variance), as all Bayesian priors are assumed to be flat
\footnote{or, equivalently, flat everywhere where the likelihood has appreciable 
power; setting upper limits based, for example, on prior observational knowledge that 
the magnetic field in the interstellar medium is less than tens of thousands of Gauss 
can yield a flat yet proper Bayesian prior, if such a property is desired. Here, 
the results are completely independent of such a choice.}. 

For simplicity, we assume that values of  $B_{\rm los}$ in the envelope follow a 
Gaussian distribution with mean equal to $B_0$ and intrinsic spread $\sigma_0$
\footnote{ Since we still have to assume a functional form for the probability 
distribution of $B_{\rm los}$ 
within the envelope, the uncertainty we derive for $B_0$ {\em is still underestimated} 
since we are not accounting for the possibility that $B_{\rm los}$ is distributed 
according to some other functional form.}. If we could observe magnetic fields with 
zero observational error, the probability to observe a value $B$ at some specific 
envelope location (the {\em likelihood} $l$ of a single observation) would be
$l = \exp\left[-(B-B_{0})^2/2\sigma_0^2\right]/
\sqrt{2\pi}\sigma_0$. Since we have observational error, complications arise. 
At any specific envelope location, there is a probability 
$\exp\left[-(B-B_{0})^2/2\sigma_0^2\right]/\sqrt{2\pi}\sigma_0$ for the magnetic 
field to have a {\em true} value $B_0$. If the (Gaussian) observational uncertainty 
at this same location is $\sigma_j$, then the probability to observe a value $B_j$ 
of the field {\em given} that its true value is $B$ is the latter probability 
multiplied by a factor $\exp\left[-(B-B_{j})^2/2\sigma_j^2\right]/\sqrt{2\pi}\sigma_j$. 
But this is not the only way we could get an observed field value $B_j$. The true 
value of the field might be $B'$, and we would still have a finite probability to 
observe $B_j$ (same reasoning as above, with $B$ substituted by $B'$). To get the 
{\em total} probability for a single observation of $B_j$, we need to integrate 
over all possible ``true'' values of the magnetic field at a single location. Doing 
this, the likelihood for a single observation $B_j$ with observational uncertainty 
$\sigma_j$ becomes
 \begin{equation}\label{like1}
l_{j}(B_0,\sigma_0) = \int_{-\infty}^{\infty} \!\!\!\!\!\!\!dB
\frac{\exp\left[-(B-B_{j})^2/(2\sigma_{j}^2)\right]}
{\sqrt{2\pi}\sigma_{j}}
\frac{ \exp\left[-(B-B_{0})^2/(2\sigma_0^2)\right]}
{\sqrt{2\pi}\sigma_0} \, .
\end{equation}
The likelihood $\mathcal{L}$ for $N$ observations of $B_j$ with individual 
gaussian uncertainties $\sigma_j$ to come from a Gaussian intrinsic probability 
distribution with mean $B_0$ and spread $\sigma_0$ is simply the product of the 
individual likelihoods, $\mathcal{L} = \prod_{j=1}^Nl_j$ which, after performing 
the integration in equation (\ref{like1}) and some algebraic manipulations, yields 
(e.g., see Venters \& Pavlidou 2007)
\begin{equation}\label{likelihood_text}
\mathcal{L} (B_0,\sigma_0)
= 
\left(
\prod_{j=1}^{N}\frac{1}{\sqrt{\sigma_0^2+\sigma_{j}^2}}
\right)
\exp\left[-\frac{1}{2} \sum_{j=1}^{N}
\frac{(B_{j} - B_{0})^2 }{\sigma_0^2+\sigma_{j}^2}
\right]\,.
\end{equation}
Note that for $\sigma_0=0$, the value of $B_0$ that maximizes equation 
(\ref{likelihood}) is exactly the weighted average, as it should be: the weighted 
average formula is the maximum-likelihood estimator of the mean when the spread 
comes purely from observational uncertainty and the observational uncertainties 
are unequal. 

The likelihood of equation (\ref{likelihood}) is a powerful tool that allows us 
to calculate the joint probability distribution of the mean $B_0$ {\em and} the 
unknown intrinsic population spread $\sigma_0$. However, if one wants to calculate 
the magnetic flux passing through a surface on which the magnetic 
field has a probability distribution of values, one finds that the flux integral 
is exactly equal to the product of the distribution mean $B_0$ times the surface 
area (in this case, modulated by $\cos \theta$ since only the line-of-sight 
component of the magnetic field is measured). Although we need to allow for 
different values of $\sigma_0$ to correctly estimate the uncertainty on $B_0$, 
the actual value of $\sigma_0$ is not of interest for the purpose of this analysis. 
We can therefore integrate it out of the likelihood (compute the {\em marginalized} 
likelihood of the mean, $\mathcal{L}_m = \int \mathcal{L} d \sigma_0$, which 
physically corresponds to calculating the probability distribution of the mean for 
{\em any} value of $\sigma_0$ allowed by the data). The marginalized likelihood 
gives the probability distribution of 
$B_0$ {\em allowing for any possible} $\sigma_0$ that can explain the data. 
The marginalized likelihood is asymmetric, and so will be the error bars. In 
addition, the location of the maximum-likelihood value of $B_0$ does not have to 
agree, and generally does not agree, with the 
estimate of the more restrictive equation (\ref{av}). 

To calculate uncertainties of the mean, we start from the peak of the marginalized 
likelihood and progressively lower the likelihood level, calculating the fractional 
integral under the curve and between the two values $B_{0-}$ and 
$B_{0+}$ left and right of the maximum-likelihood value, which are such that 
$\mathcal{L}_m(B_{0-}) = \mathcal{L}_m(B_{0+})$. We do so until we find the 
$\mathcal{L}_m$ value for which this fractional integral is equal to $68.26\%$. 
The $+$ and $-$ $1\sigma$ uncertainties are 
then defined as $B_{0+} - B_{\rm max\mathcal{L}}$ and $B_{\rm max 
\mathcal{L}}- B_{0-}$, respectively. 

Upper limits can also be calculated starting with the marginalied likelihood, 
as follows. The $2\sigma$ upper limit is that value of 
$|B_0|$ for which the area under the marginalized likelihood curve between 
$-|B_0|$ and $|B_0|$ is $95.4\%$. In other words, a fractional area equivalent to 
the gaussian $2\sigma$ is included between $-|B_0|$ and $|B_0|$. Note that, 
because of the different nature of the question (in this 
case we are interested in fixed values of $B_0$ rather than equal $\mathcal{L}_m$ 
likelihood levels), this is {\em not} simply the maximum-likelihood value of 
$B_0$ plus two times the error. In addition, the marginalized likelihood has much 
longer tails than a gaussian, and hence $1\sigma$, $2\sigma, ...$ values do not 
scale linearly (if we desire to keep the terms representing the same fractional 
integrals as in the case of a gaussian). 

\begin{acknowledgements}
Invaluable discussions and comments by Vasiliki Pavlidou are gratefully acknowledged. 
We thank Mark Heyer, Robert Dickman and Dan Marrone for educating us on a number of
observational issues, for other science discussions, and for their critical comments. 
TM acknowledges partial support by the National Science Foundation under grant 
AST-07-09206. KT  acknowledges support by NSF grants AST 02-06216 and AST02-39759,  
by the NASA Theoretical Astrophysics Program grant NNG04G178G and  by the Kavli 
Institute for Cosmological Physics at the University of Chicago through grants 
NSF PHY-0114422 and NSF PHY-0551142 and an endowment from the Kavli Foundation 
and its founder Fred Kavli.
\end{acknowledgements}

\end{document}